\documentclass[aps,pre,twocolumn,amsmath,amssymb,superscriptaddress,floatfix]{revtex4}
\usepackage{graphicx}
\usepackage{bm}
\bibstyle{apsrev.bib}

\newcommand{\be}{\begin{equation}}
\newcommand{\ee}{\end{equation}}
\newcommand{\beqn}{\begin{eqnarray}}
\newcommand{\eeqn}{\end{eqnarray}}

\begin{document}

\title{Numerical study of the critical behavior of the Ashkin-Teller model at a line defect}

\author{P\'eter Lajk\'o}
 \email{peter.lajko@ku.edu.kw}
 \affiliation{Department of Physics, Kuwait University, P.O. Box 5969, Safat 13060, Kuwait}
%
%
\author{Ferenc Igl\'oi}%
 \email{igloi@szfki.hu}
 \affiliation{Research Institute for Solid State Physics and Optics,
H-1525 Budapest, P.O.Box 49, Hungary}
 \affiliation{Institute of Theoretical Physics,
Szeged University, H-6720 Szeged, Hungary}

\date{\today}

\begin{abstract}
We consider the Ashkin-Teller model on the square lattice, which is
represented by two Ising models ($\sigma$ and $\tau$) having a
four-spin coupling of strength, $\epsilon$, between them.  We
introduce an asymmetric defect line in the system along which the
couplings in the $\sigma$ Ising model are modified.  In the
Hamiltonian version of the model we study the scaling behavior of the
critical magnetization at the defect, both for $\sigma$ and for $\tau$
spins by density matrix renormalization. For $\epsilon>0$ we observe
identical scaling for $\sigma$ and $\tau$ spins, whereas for
$\epsilon<0$ one model becomes locally ordered and the other locally
disordered.  This is different of the critical behavior of the
uncoupled model ($\epsilon=0$) and is in contradiction with the
results of recent field-theoretical calculations.
\end{abstract}


\maketitle

\section{Introduction}
\label{sec:intro}
In a system which is divided into two parts by a defect plane
translational invariance is broken and the physical properties are
different in the defect region, which has a width of the correlation
length, $\xi$. At the critical point where $\xi$ is divergent the
scaling properties of local quantities, such as the defect
magnetization or the spin-spin correlation function, could be
different from that in the bulk\cite{ipt93}. Relevance or irrelevance
of the perturbation caused by a weak defect can be analyzed within the
frame of phenomenological scaling
theory\cite{burkhardt81_rev,burkhardt81,diehl_diet_eisenr,eisenr_burkh}. If
the perturbation is coupled to the local energy operator, then the
bulk fixed point is stable if the correlation length critical exponent
of the pure system, $\nu$, is greater than $1$. In this case the
defect exponents are the same as at any other point of the bulk. In
the opposite case for $\nu < 1$, generally a new fixed point governs
the local critical behavior, the properties of which are different for
weakened and for enhanced local couplings. For weakened defect
couplings the defect usually renormalizes to a cut and the local
critical behavior is the same as at the ordinary surface critical
point\cite{binder83,diehl86,pleimling04}. On the contrary for enhanced
defect couplings the defect usually renormalizes to an ordered
interface and the local critical exponents are the same as at the
extraordinary surface transition\cite{binder83,diehl86,pleimling04}.
Examples for modified defect critical behavior can be found in the
two-dimensional (2D) $q=3$ state Potts model\cite{nb82,it93}, in the
Baxter-Wu model\cite{bti06} or in the Ashkin-Teller (AT)
model\cite{lti07}.

According to scaling considerations in a system having a correlation
length exponent $\nu=1$ a defect is a marginal perturbation. This
happens for the 2D Ising model for which the local
magnetization\cite{bariev79} and the spin-spin correlation function at
the defect\cite{McC_Perk} has been exactly calculated and the local
critical exponent of the magnetization is found to be a continuous
function of the strength of the defect.  These results have been
generalized recently for inhomogeneous line defects, in which the
defect coupling is a smooth function of the position\cite{naon_trobo}.

In a system which has more complicated interaction between the spin
variables one can define special defects which influence only some
parts of the interaction. This type of asymmetric defect has been
considered recently by Naon\cite{naon} in the AT\cite{AT} and in the
Baxter models\cite{baxter}. We remind that both the AT and the Baxter
models are expressed in terms of two sets of Ising spin
variables\cite{fan72}, say $\sigma$ and $\tau$, with two-site ($K_2$)
and four-site ($K_4$) interactions. The asymmetric defect introduced
by Naon is represented by a line of modified two-spin couplings in one
of the Ising models (say for the $\sigma$ spins), whereas the other
interactions are left unchanged. Using field-theoretical methods Naon
has determined the decay of the critical spin-spin correlation
function, both for the $\sigma$ and for the $\tau$ spins and the
calculated local magnetization exponents are found to be independent
of the interaction between the Ising models, which is measured by
$\epsilon=K_4/K_2$. Similar results are obtained in the case of an
asymmetric and inhomogeneous line defect in the AT
model\cite{naon_trobo}. These results are somewhat surprising, since
in the pure AT and Baxter models $\epsilon$ is a marginal perturbation
and several (bulk and surface) critical exponents are continuous
functions\cite{baxter_book} of $\epsilon$.

In this paper we are going to revisit this type of problem and study
numerically the local critical behavior at an asymmetric defect in the
AT model. Our aim is to shed some light to the physical mechanism
which is behind this problem and to confront the measured critical
exponents with the results of field-theoretical calculations.  Here we
shall consider both chain and ladder type defects\cite{ipt93} and the
system is studied by the transfer matrix, which is perpendicular to
the defect. We use the Hamiltonian limit\cite{KdNK81,IS84} of the
transfer matrix and calculate the local magnetization both for
$\sigma$ and $\tau$ spins by density matrix renormalization\cite{DMRG}
(DMRG). At the critical point the local magnetization exponents are
then deduced through finite-size scaling.

The structure of the paper is the following. The model and its basic
properties are described in Sec.~\ref{sec:AT}.  Numerical results of
the scaling of the interface magnetization are presented in
Sec.~\ref{sec:num} and discussed in Sec.~\ref{sec:disc}.

\section{The AT model with a defect line}
\label{sec:AT}
The AT model is defined in terms of a four-state spin
variable\cite{AT}, which is represented by a pair of Ising spins:
$\sigma_i=\pm 1$ and $\tau_i=\pm 1$ at the lattice site $i$.  Between
the same set of spins there is nearest neighbor interaction of
strength $K_2$ and the two Ising models are coupled by the product of
the energy densities, which is given by a four-spin term: $
K_4\sigma_i \sigma_j \tau_i \tau_j$, where $i$ and $j$ are nearest
neighbors. We consider the system on the square lattice and work with
the row-to-row transfer matrix ${\cal T}_{AT}$.  In the Hamiltonian
limit the transfer matrix can be written as ${\cal T}_{AT} \sim
\exp(-\kappa{\cal H}_{AT})$, where ${\kappa}$ is the lattice spacing
in the ``time" direction and ${\cal H}_{AT}$ is a one-dimensional
quantum Hamiltonian given by\cite{KdNK81,IS84}
\beqn
{\cal H}_{AT}=&-&\sum_i(\sigma_i^z \sigma_{i+1}^z + \tau_i^z \tau_{i+1}^z)
-h \sum_i (\sigma_i^x+\tau_i^x)\cr
&-&\epsilon\left[\sum_i\sigma_i^z \sigma_{i+1}^z\tau_i^z \tau_{i+1}^z
+h \sum_i \sigma_i^x \tau_i^x \right]\,.
\label{H_AT}
\eeqn
Here $\sigma_i^{x,z}$ and $\tau_i^{x,z}$ are two sets of Pauli
matrices at site $i$ and $h$ is the strength of the transverse field,
which plays the role of the temperature in the classical system. As
before the ratio of the couplings is denoted by $\epsilon=K_4/K_2$.

The system in Eq.(\ref{H_AT}) is self-dual and the self-duality line:
$h_c=1$ represents the critical line separating the ferromagnetic and
the paramagnetic phases of the system for $-1/\sqrt{2} \le \epsilon
\le 1$. In the region $-1<\epsilon \le -1/\sqrt{2}$, there is a
so-called ``critical fan", which extends to both sides of the
self-duality line and in which the system stays
critical~\cite{KdNK81}.

The critical properties of the AT model are exactly known through
conformal invariance and Coulomb-gas
mapping~\cite{cardy87,nienhuis87}. The decay of the spin-spin
correlations:
\be
\langle 0 | \sigma_i^{z} \sigma_{i+r}^{z}|0\rangle= \langle 0 | \tau_i^{z} \tau_{i+r}^{z}|0\rangle \sim r^{-2x_m}
\ee
is characterized by the anomalous dimension, $x_m=1/8$, which does not
depend on the value of the coupling $\epsilon$. On the contrary the
decay of the connected energy-energy correlations:
\be
\langle 0 | \sigma_i^{x} \sigma_{i+r}^{x}|0\rangle-\langle 0 | \sigma_i^{x}|0\rangle \langle 0 |\sigma_{i+r}^{x}|0\rangle \sim r^{-2x_e}
\ee
involves the anomalous dimension: $x_e=\pi/[2 \arccos(-\epsilon)]$,
which is $\epsilon$ dependent\cite{KdNK81}. Similar holds for the
correlation length critical exponent $\nu$, which is given by:
$\nu=1/(2-x_e)$ for $-1/\sqrt{2} \le \epsilon \le 1$ whereas it is
formally infinite in the critical fan.

Finally, the decay of the end-to-end correlation function at the
critical point involves the surface magnetization scaling dimension,
$x_m^s$:
\be
\langle 0 | \sigma_{-L}^{z} \sigma_{L}^{z}|0\rangle= \langle 0 | \tau_{-L}^{z} \tau_{L}^{z}|0\rangle \sim L^{-2x_m^s}
\label{end_end}
\ee
which is also coupling dependent\cite{vGR87}: $x_m^s=\arccos(-\epsilon)/\pi$.

The marginal operator for the AT model is associated with the
four-spin term, $\sigma^z_i\sigma^z_{i+1}\tau^z_i\tau^z_{i+1}$, which
has a scaling dimension, $x_4=2=D$, independently of $\epsilon$, for
$-1 \le \epsilon \le 1$.

\subsection{Ladder and chain defects}

A line defect in the 2D classical model is put in the ''time``
direction and can be of two types: chain defect or ladder defect (see
Fig.~6.1 of Ref.~\cite{ipt93}). In the problem studied by
Naon\cite{naon} and we consider here a line of two-spin couplings
between the $\sigma$-spins are modified. In the Hamiltonian limit,
when the defect is placed between sites $i=0$ and $i=1$ (ladder
defect) the perturbation is given by:
\be
{\cal V}_{\rm ladder}=-(J-1) \sigma_0^z \sigma_1^z\;,
\label{ladder}
\ee
where $J$ is the strength of the defect. On the other hand for a chain
defect with modified two-spin couplings at line $i=0$ the perturbation
in the Hamiltonian limit involves the term:
\be
{\cal V}_{\rm chain}=-(\widetilde{h}-h) \sigma_0^x\;.
\label{chain}
\ee
The ladder and chain defects transforms into each other through
duality\cite{lti07} and their strengths are related as: $\widetilde{h}
\leftrightarrow 1/J$.

For two decoupled Ising models with $\epsilon=0$ the local
magnetization exponents at the defect site, $i=0$, generally are
different for the $\sigma$ and for the $\tau$ spins, which are denoted
by $x_m^{\sigma}$ and $x_m^{\tau}$, respectively. While $x_m^{\tau}$
keeps its bulk value, $x_m^{\sigma}$ is a continuous function of the
strength of the defect\cite{bariev79,McC_Perk}:
\be
x_m^{\sigma}(J)=\frac{2}{\pi^2} \arctan^2(1/J),\quad x_m^{\tau}=1/8,\quad \epsilon=0\,,
\label{marg_0}
\ee
for a ladder defect and
\be
x_m^{\sigma}(\widetilde{h})=\frac{2}{\pi^2} \arctan^2(\widetilde{h}),\quad x_m^{\tau}=1/8,\quad \epsilon=0\,,
\label{marg_1}
\ee
for a chain defect. For $\sigma$ spins the marginal operator is the
local energy density, which has its anomalous dimension
$x_e^{\sigma}=1$, independently of the value of $J$ or
$\widetilde{h}$.

If we switch on the interaction between the Ising models the defect
critical behavior could be modified.  For small $\epsilon$ the
anomalous dimension of the four-spin operator is
$x_4=x_e^{\sigma}+x_e^{\tau}=2$, which is just the marginal value. We
note, however, that field-theoretical calculations of the critical
spin-spin correlation function\cite{naon} come to the conclusion, that
$\epsilon$ is an irrelevant variable. Here we revisit this problem by
numerical methods.

\section{Numerical study}
\label{sec:num}
Here we consider the Hamiltonian version of the AT model at a finite
lattice with $-L \le i \le L$ and use fixed-spin boundary conditions:
$\sigma_{\pm L}^z=\tau_{\pm L}^z=+1$. Using the DMRG method we
calculate the ground-state expectation value of the magnetization
operators at the defect, $\langle 0 | \sigma_0^x|0 \rangle=
m_0^{\sigma}(L)$ and $\langle 0 | \tau_0^x|0
\rangle=m_0^{\tau}(L)$. The parameters in Eq.(\ref{H_AT}) are taken at
different points of the critical line: $h=h_c=1$ and $-1 < \epsilon
\le 1$ and we have considered different strengths of the defect:
$\widetilde{h}>0$ and $J>0$.  According to finite-size scaling theory
these magnetizations asymptotically behave as:
\be
m_0^{\alpha}(L) \sim L^{-x_m^{\alpha}},\quad \alpha=\sigma,\tau\;.
\label{fss}
\ee
In the numerical calculation we went up to $L=85$ and in the DMRG
method we have generally kept around $m=150$ states of the density
matrix in order to obtain a good numerical accuracy.

From the values of the matrix element in Eq.(\ref{fss}) at two
different sizes, $L$ and $b L$, we have calculated effective,
size-dependent exponents through two-point fits:
\be
\frac{\ln m_{0}^{\alpha}(bL)-\ln m_{0}^{\alpha}(L)}{\ln b}=x_{m}^{\alpha}(L)\,.
\label{fs_exp}
\ee
In order to obtain the same numerical accuracy for the different
lengths, we keep the ratio $b$ between neighboring sizes approximately
constant. The effective exponents evolve towards their exact values
when the mean size associated with the two-point fit, $\langle
L\rangle=L(b+1)/2$, tends to infinity.

In the actual calculation we consider first in Sec.\ref{sec:num_lim}
the known limiting cases (decoupling limit, system without defect) in
order to check the accuracy of the numerical method. Afterward we
study the general model with defect for $\epsilon>0$ and $\epsilon<$
in Sec.\ref{sec:num_pos} and in Sec.\ref{sec:num_neg}, respectively.

The numerical calculations are performed on several PC-s, the total CPU
time being equivalent to $\sim 250$ days in a single 2.4 GHz processor.
We note that an alternative numerical method is to use Monte Carlo (MC)
simulations on the classical model with $L \times M$ sites. As can be
seen in an investigation of a related problem\cite{cissz} the DMRG and the
MC methods generally provide results with approximately the same accuracy
using the same CPU times.

\subsection{Decoupled or non-defected systems}
\label{sec:num_lim}

\begin{figure}
\begin{center}
\includegraphics[width=8cm]{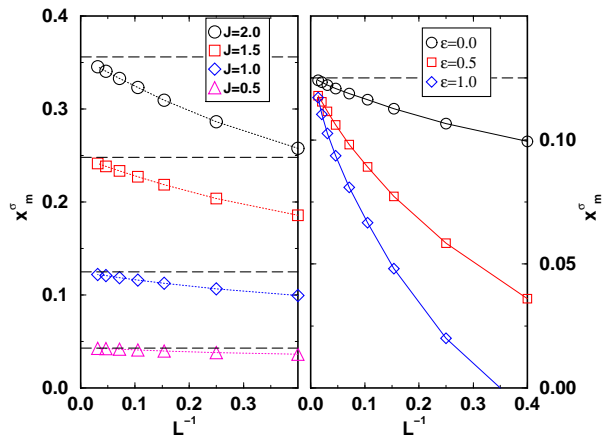}
\end{center}
\caption{Numerical estimates of the magnetization exponent
  $x_m^{\sigma}(L)$ for different finite systems of length $L$ at
  special points of the phase diagram. Left panel: at the decoupling
  point, $\epsilon=0$, for various values of the strength of the
  ladder defect, $J$. Right panel: non-defected AT model for different
  values of the coupling, $\epsilon$. The exact results are indicated
  by horizontal broken lines.}
\label{fig:1}
\end{figure}


We start with the decoupled model and calculate the critical local
magnetization of the $\sigma$ Ising model having a ladder defect. The
finite-size magnetization exponents, $x_m^{\sigma}(L)$, which have
been obtained through Eq.(\ref{fs_exp}) are plotted as a function of
$1/L$ in the left panel of Fig.\ref{fig:1} for different strength of
the defect, $J$. Extrapolating for large $L$ the limiting exponents
are in good agreement with the analytical results in
Eq.(\ref{marg_0}), which are indicated by dashed lines.

We have also studied the AT model without defects, but for different
values of the coupling, $\epsilon \ge 0$. The finite-size
magnetization exponents, which in this case correspond to the bulk
exponents, are plotted as a function of $1/L$ in the right panel of
Fig.\ref{fig:1}. As seen in the figure the extrapolated exponents are
fairly close to the exact value, $x_m=1/8$, which indeed does not
depend on $\epsilon$. The correction terms, however, are larger for a
more strongly coupled system.

\subsection{Positive coupling: $\epsilon>0$}
\label{sec:num_pos}
\begin{figure}
\begin{center}
\includegraphics[width=8cm]{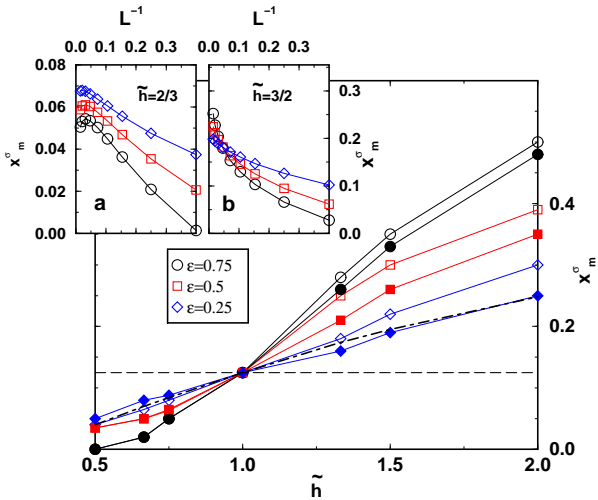}
\end{center}
\caption{Finite-size magnetization exponents $x_m^{\alpha}(L)$
  calculated at the largest length $L$ at a chain defect of strength
  $\widetilde{h}$ and for positive values of the coupling,
  $\epsilon>0$. The open (filled) symbols are for $\alpha=\sigma$
  ($\alpha=\tau$) spins.  Note, the very similar scaling behavior of
  $\sigma$ and $\tau$ spins.  The exact results for the decoupled
  model with $\epsilon=0$ are indicated by dotted and broken lines,
  for $\sigma$ and $\tau$ spins, respectively. These represent the
  field-theoretical result\cite{naon}.  Insets: Size-dependence of the
  effective magnetization exponents for $\sigma$ spins. a) For
  enhanced defect strength, $\widetilde{h}=2/3$, the finite-size
  exponents are expected to approach $x_m^{\alpha}(L)=0$, which
  corresponds to an EI fixed point. b) For reduced defect strength,
  $\widetilde{h}=3/2$, the finite-size exponents are expected to
  approach $x_m^{\sigma}=x_m^{\tau}=x_m^s$, which corresponds to an OI
  fixed point.}
\label{fig:2}
\end{figure}

This part of the phase diagram of the Hamiltonian model corresponds to
the classical model with positive Boltzmann-weights, thus the results
obtained in this domain should be compared with the field-theoretical
calculations in Ref.\cite{naon}. First we present in Fig.\ref{fig:2}
the finite-size magnetization exponents (both for $\sigma$ and $\tau$
spins) which are calculated at the largest length, $L=85$. Here we
have a system with a chain defect of strength,
$\widetilde{h}=1/2,~2/3,~3/4,~1.,~4/3,~3/2$ and $2$ and various
positive values of the coupling.  For the sake of comparison we also
present the analytical results in the decoupling limit, $\epsilon=0$,
which are plotted with dotted-dashed and broken lines for $\sigma$ and
$\tau$ spins, respectively. We note that according to
field-theoretical investigations\cite{naon} these results should hold
for $\epsilon>0$, too. The numerical results in Fig.\ref{fig:2} seem
to be in contradiction with the field-theoretical conjectures in two
respects. i) The effective ($L$-dependent) magnetization exponents
vary with the coupling, $\epsilon>0$, in particular for large
$\epsilon$ there is a considerable difference from the value at
$\epsilon=0$. ii) The magnetization exponents at the $\tau$ spins are
different from the predicted bulk value, $x_m=1/8$, and these are
close to that measured values at $\sigma$ spins at the same system.
The trend of the effective exponents with the size of the system is
different for enhanced ($J>1$ or $\widetilde{h}<1$) and reduced ($J<1$
or $\widetilde{h}>1$) defect couplings, respectively.

\underline{For enhanced defect couplings} the local magnetization
exponents are smaller than the values at the decoupling limit and the
finite-size exponents - for sufficiently large sizes - are decreasing
with $L$. This is illustrated in the inset a) of Fig.\ref{fig:2} for
$\widetilde{h}=2/3$. We expect that this decreasing tendency will
continue for larger sizes and the extrapolated exponents will approach
$x_m^{\sigma}=x_m^{\tau}=0$.  This type of behavior is characteristic
for an ordered defect (see Eqs.(\ref{marg_0}) and (\ref{marg_1}) for
$J \to \infty$ and $\widetilde{h} \to 0$, respectively), and we assume
that the local transition for enhanced couplings is governed by an
extraordinary interface (EI) fixed point.

\underline{For reduced defect couplings} the effective, $L$-dependent
exponents are larger than their value at the decoupling point and
these are continuously increasing with $L$. In order to check a
possible overshooting effect we have made calculations up to $L \sim
500$, but no bending of the curves are found. We have also compared
the magnetization exponents at the two sublattices,
i.e. $x_m^{\sigma}$ with $x_m^{\tau}$.  For a finite $L$,
$x_m^{\sigma}(L)$ is larger than $x_m^{\tau}(L)$, however the limiting
value of the two exponents are very close to each other, see
Fig.\ref{fig:2} for $\widetilde{h}>1$. We expect, that for the true
exponents we have $x_m^{\sigma}=x_m^{\tau}$. A more difficult point is
to decide about the eventual $\widetilde{h}$-dependence of the
exponents.  Here one should note that the effective exponents show a
quite strong $\widetilde{h}$, as well as $L$-dependence.  As
illustrated in the inset b) of Fig.\ref{fig:2} the extrapolation of
the data with $L$ is very difficult, at least for not too large values
of $\widetilde{h}$. To interpret the data we use the hypothesis, that
the interface critical behavior is controlled by an ordinary interface
(OI) fixed point. In this scenario the true exponents are independent
of $\widetilde{h}$ and their values are the same as at a free surface,
as given below Eq.(\ref{end_end}). Indeed, for larger values of
$\widetilde{h}$ the extrapolated exponents are fairly close to
$x_m^s(\epsilon)$.

\subsection{Negative coupling: $\epsilon<0$}
\label{sec:num_neg}

\begin{figure}
\begin{center}
\includegraphics[width=8cm]{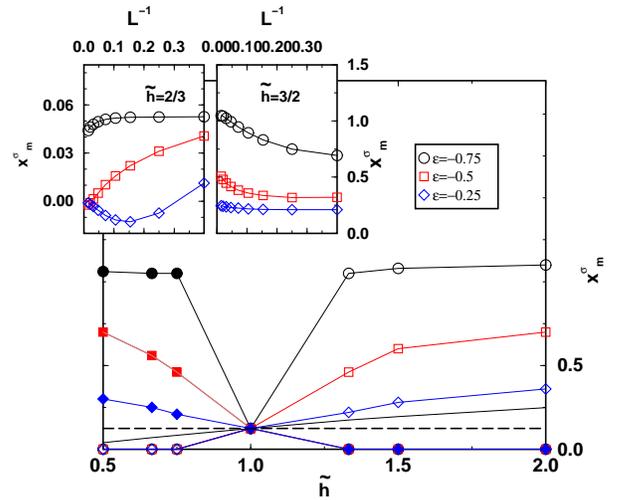}
\end{center}
\caption{The same as in Fig.\ref{fig:2} but for negative coupling,
  $\epsilon<0$. Note, the complementary scaling behavior of $\sigma$
  and $\tau$ spins: $x_m^{\sigma}(\widetilde{h})\approx
  x_m^{\tau}(1/\widetilde{h})$.  Insets: Size-dependence of the
  effective magnetization exponents for $\sigma$ spins. a) For
  enhanced defect strength, $\widetilde{h}=2/3$, the finite-size
  exponents are expected to approach $x_m^{\alpha}(L)=0$, which
  corresponds to an EI fixed point. b) For reduced defect strength,
  $\widetilde{h}=3/2$, the finite-size exponents are expected to
  approach a non-trivial value: $x_m^{\sigma}(\epsilon)$, which
  corresponds to a special OI fixed point.}
\label{fig:3}
\end{figure}

This part of the phase diagram of the Hamiltonian model corresponds to
the classical model with negative Boltzmann-weights. The calculated
finite-size magnetization exponents for the same absolute values of
$\epsilon$ and for the same values of $\widetilde{h}$ as in
Sec.\ref{sec:num_pos} are shown in Fig.\ref{fig:3}. The two figures,
Fig.\ref{fig:3} and Fig.\ref{fig:2}, show several differences. The
most important difference is that here the limiting values of
$x_m^{\sigma}$ and $x_m^{\tau}$ are different. The effective exponents
for $\sigma$ spins with $\widetilde{h}$ are approximately equal to the
effective exponents for $\tau$ spins, however at $1/\widetilde{h}$. As
far as the defect exponents on the $\sigma$ spins are concerned the
trend is similar as for $\epsilon>0$.  For enhanced defect couplings
the finite-size exponents seem to approach $x_m^{\sigma}=0$, thus the
$\sigma$ spins are locally ordered and their critical behavior at the
defect is controlled by the EI fixed point. On the contrary for
reduced couplings at the defect the finite-size exponents are
monotonously increasing with the size and they seem to approach a
non-trivial value: $x_m^{\sigma}=x_{\rm def}(\epsilon)$.  We expect,
that this limiting value depends only on $\epsilon$, but does not
depend on $J<1$ or $\widetilde{h}>1$, thus it can be identified as a
special interface exponent in the model with a vanishing bond, $J=0$.

\section{Discussion}
\label{sec:disc}
\begin{figure}
\begin{center}
\vskip 1cm
\includegraphics[width=8cm]{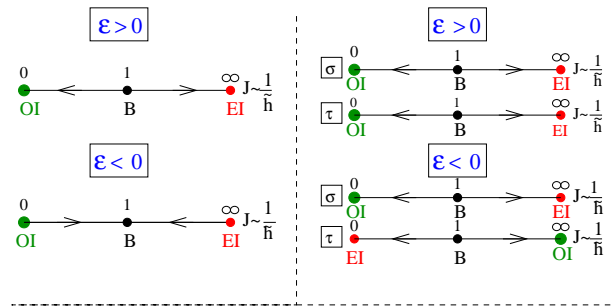}
\end{center}
\caption{Schematic RG phase diagram at a defect line of strength $J
  \sim 1/\widetilde{h}$ in the critical AT model. Left panel:
  symmetric defect, right panel: asymmetric defect for the $\sigma$
  and $\tau$ spins.  The RG flow is different for different signs of
  the bulk four-spin coupling $\epsilon$. For $\epsilon>0$, the flow
  is towards the ordinary interface fixed point (OI) when $J<1$ and
  the extraordinary interface fixed point (EI) when $J>1$, both for
  symmetric and asymmetric defects. When $\epsilon<0$, the flow is to
  the bulk fixed point for a symmetric defect. For asymmetric defect
  the RG flow is different for $\sigma$ and $\tau$ spins.}
\label{fig:4}
\end{figure}

The numerical results presented in the previous Section about the
interface critical behavior of the AT model with a (asymmetric) line
defect can be interpreted in terms of an RG phase phase diagram, which
is shown in the right panel of Fig.\ref{fig:4} both for the $\sigma$
and the $\tau$ spins.  As a comparison in the left panel of
Fig.\ref{fig:4} we show the RG phase diagram for such a (symmetric)
defect, which is proportional with the local energy density and given
by:
\be
\widetilde{\cal V}_{\rm chain}=-(\widetilde{h}-h)(\sigma_0^x + \tau_0^x +
\epsilon \sigma_0^x \tau_0^x)\;,
\label{V_h}
\ee
for a chain defect and similarly for a ladder defect. In this case the
$\sigma$ and $\tau$ spins play equivalent role and the interface
critical behavior has been studied previously in
Refs.\cite{bti06,lti07}. According to these results, which are in
agreement with a relevance-irrelevance
analysis\cite{burkhardt81_rev,burkhardt81,diehl_diet_eisenr,eisenr_burkh},
the bulk fixed point (B) in the system is unstable for
$\epsilon>0$. The numerical results also indicate that the critical
behavior at the defect is governed by the OI and EI fixed points, for
reduced and enhanced strength of the defect, respectively. This type
of RG phase diagram is suggested to be valid for the asymmetric defect
too, as illustrated in the right panel of Fig.\ref{fig:4}.  This
result indicate, that the phase-diagram in the left panel of
Fig.\ref{fig:4} for $\epsilon>0$ is probably valid in a coarse-grained
description: the direction of the flow at B depends only on the
condition, if the local energy at the defect is reduced or enhanced
with respect to the pure system.


For negative coupling, $\epsilon<0$, the structure of the
phase-diagrams are changed, both for symmetric and asymmetric
defects. For symmetric defects the bulk fixed-point becomes the stable
one\cite{bti06,lti07}. It is certainly understandable, that the EI
fixed point is unstable, since for $\epsilon<0$ ordered defects would
bring a positive contribution to the energy. The same reasoning holds
for the asymmetric defect, too. Thus for $\epsilon<0$ the $\sigma$ and
the $\tau$ spins can not be ordered at the same time. In this case in
the energetically favorable state one set of spins is disordered
(being in the OI fixed point) and the other set of spins is ordered
(EI fixed point). This is the situation we have found numerically and
which is shown in the phase diagram in the right panel of
Fig.\ref{fig:4}.

The field-theoretical results about the spin-spin correlation function
at the defect line suggest a different RG phase
diagram\cite{naon}. According to this theory, which expected to hold
for positive Boltzmann-weights ($\epsilon>0$) the coupling term is an
irrelevant perturbation and the critical behavior at the defect is the
same as for the decoupled systems. Our numerical results are in
contradiction with this scenario. In this respect we mention that in
the asymmetric defect problem two potentially marginal operators are
involved. One is the product of the energy-densities in the AT model
and the second is the local energy-density (line defect) in the Ising
model with $\sigma$ spins. The interplay of these two perturbations at
the critical point could result in the phase diagram in
Fig.\ref{fig:4} which is consistent with our numerical calculations.

\begin{acknowledgments}
This work has been supported by Kuwait University Research
Grant No. SP01/10. We thank L. Turban for useful discussions.
\end{acknowledgments}

\end{document}